\documentclass[sigconf]{acmart}

\AtBeginDocument{%
  }

\setcopyright{acmlicensed}
\acmConference[CIKM '26]{The 35th ACM International Conference on Information and Knowledge Management}{November 7--11, 2026}{Rome, Italy}

\usepackage{multirow}
\usepackage{amsmath}
\usepackage{xcolor}
\usepackage{balance} 

\copyrightyear{2026}
\acmYear{2026}
\setcopyright{cc}
\setcctype{by-nc-nd}
\acmConference[CIKM '26]{Proceedings of the 35th ACM International Conference on Information and Knowledge Management}{November 07--11, 2026}{Rome, Italy}
\acmBooktitle{Proceedings of the 35th ACM International Conference on Information and Knowledge Management (CIKM '26), November 07--11, 2026, Rome, Italy}
\acmDOI{10.1145/3799682.3839876}
\acmISBN{979-8-4007-2539-5/2026/11}

\begin{document}

\title[Personalized Task Dependency Graphs for Mitigating Signal Erosion in Multi-Task Recommendation]{{Personalized Task Dependency Graphs for Mitigating Signal Erosion in Multi-Task Recommendation}}

\author{Fuyuan Liu}
\email{liufuyuan2@huawei.com}
\orcid{0009-0001-5008-0876}
\authornote{Both authors contributed equally to this research (co-first authors).}
\affiliation{%
  \institution{Huawei Technologies Co., Ltd.}
  \city{Shanghai}
  \country{China}
}

\author{Tiandeng Wu}
\email{wutiandeng1@huawei.com}
\authornotemark[1]
\authornote{Corresponding author.}
\affiliation{%
  \institution{Huawei Technologies Co., Ltd.}
  \city{Shanghai}
  \country{China}
}

\author{Yaqun Fang}
\email{fangyaqun@huawei.com}
\affiliation{%
  \institution{Huawei Technologies Co., Ltd.}
  \city{Dongguan}
  \state{Guangdong}
  \country{China}
}

\author{Wei Zhou}
\email{zhouwei281@huawei.com}
\affiliation{%
  \institution{Huawei Technologies Co., Ltd.}
  \city{Nanjing}
  \state{Jiangsu}
  \country{China}
}

\author{Zehao Zhou}
\email{zhouzehao@huawei.com}
\affiliation{%
  \institution{Huawei Technologies Co., Ltd.}
  \city{Shanghai}
  \country{China}
}

\author{Wenping Chen}
\email{chenwenping15@huawei.com}
\affiliation{%
  \institution{Huawei Technologies Co., Ltd.}
  \city{Shanghai}
  \country{China}
}

\author{Qishun Mei}
\email{meiqishun1@huawei.com}
\affiliation{%
  \institution{Huawei Technologies Co., Ltd.}
  \city{Shanghai}
  \country{China}
}

\author{Jiaxin Zhou}
\email{zhoujiaxin15@huawei.com}
\affiliation{%
  \institution{Huawei Technologies Co., Ltd.}
  \city{Shanghai}
  \country{China}
}

\author{Heng Chang}
\email{changh17@tsinghua.org.cn}
\affiliation{%
  \institution{Huawei Technologies Co., Ltd.}
  \city{Beijing}
  \country{China}
}

\author{Yi Cao}
\email{caoyi23@huawei.com}
\affiliation{%
  \institution{Huawei Technologies}
  \city{Shanghai}
  \country{China}
}

\author{Jiandong Ding}
\email{dingjiandong2@huawei.com}
\affiliation{%
  \institution{Huawei Technologies}
  \city{Shanghai}
  \country{China}
}
\renewcommand{\shortauthors}{Fuyuan Liu et al.}
\begin{abstract}
Optimizing multiple conversion objectives is a core challenge in industrial recommendation, often limited by signal erosion in rigid architectures. Existing Multi-Task Learning (MTL) methods typically enforce uniform dependency strengths across a static conversion funnel, overlooking how task correlations naturally vary based on item characteristics. Hierarchical message passing along these fixed chains leads to cumulative signal attenuation, which degrades performance on sparse, deep-funnel objectives.
To address this, we propose the Personalized Task Dependency Graphs (\textbf{PTDG}). While respecting necessary physical causal constraints (e.g., \textit{Click} $\to$ \textit{Pay}), PTDG dynamically ``rewires'' the \textit{intensity} of dependency pathways for each item via low-rank approximation to ensure structural robustness. We implement a GCN-based propagation with hard causal masking to establish adaptive information shortcuts. Additionally, we introduce an Adaptive Progressive Masking (APM) strategy that decouples shared parameters according to task sparsity, helping to stabilize optimization. Experiments on KuaiRand1K and an industrial dataset show that PTDG significantly improves AUC on sparse conversion tasks by up to 1.45\%, while maintaining comparable performance on dense objectives. Online A/B testing shows PTDG improves Conversion Rate (CVR) by 1.2\% and effective
Cost Per Mille (eCPM) by 1.9\% relative to the baseline.

\end{abstract}



\maketitle

\section{Introduction}

\begin{figure}[htbp]
  \centering
  \includegraphics[width=\linewidth]{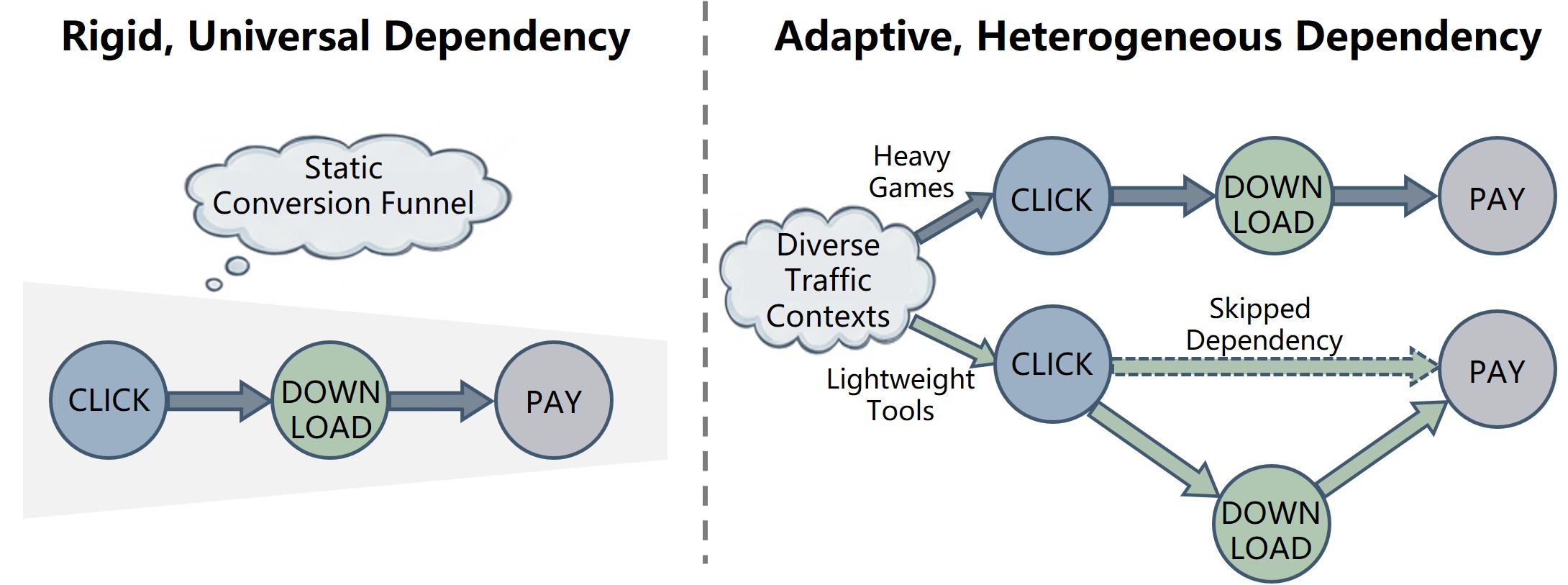}
  \caption{Contrast between rigid funnels (left) with uniform dependency and our personalized dependency graph (right) where edge intensities adapt per instance.}
  \label{background}
\end{figure}
\begin{figure*}[t]
    \centering
    \includegraphics[width=0.95\linewidth]{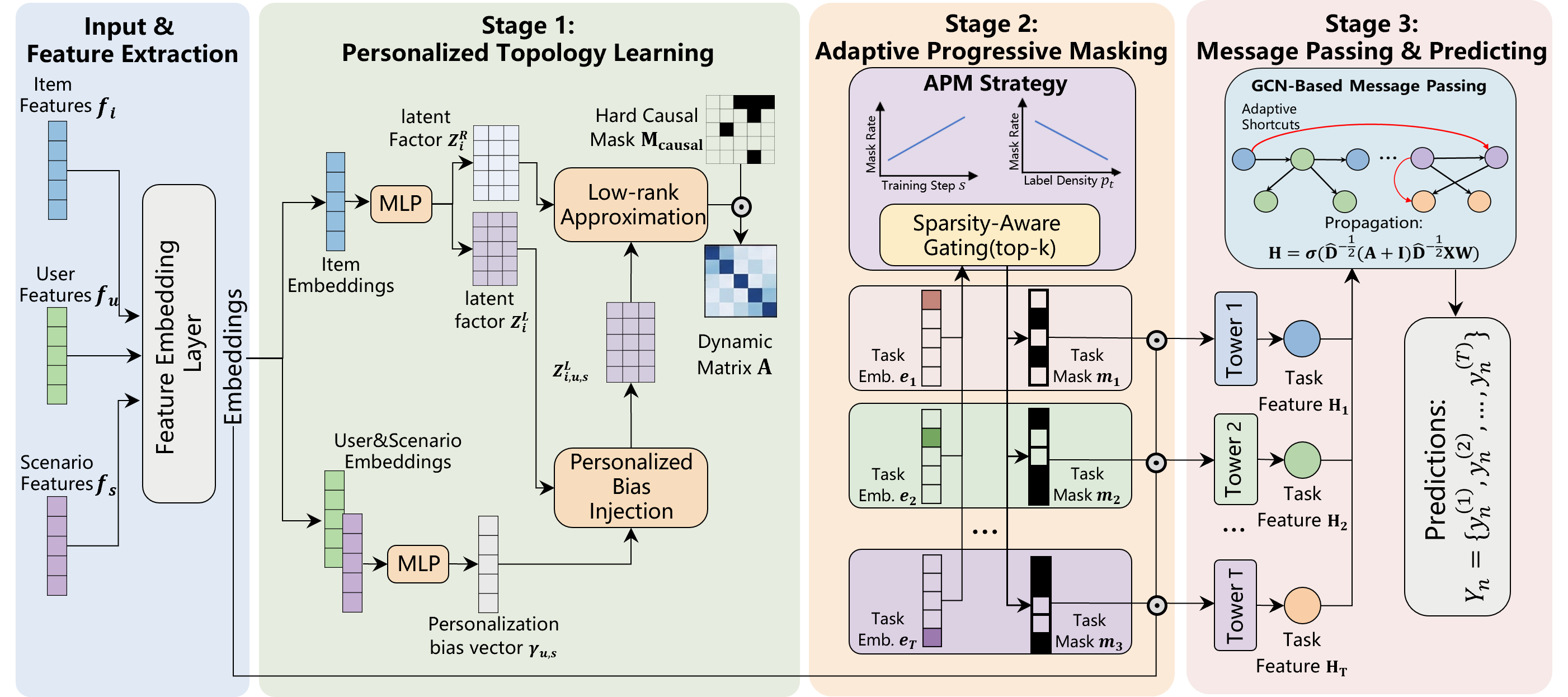}
    \caption{Overall structure of PTDG. The approach can be divided into three stages: personalized topology learning, adaptive progressive masking and message passing \& predicting.}
    \label{structure}
\end{figure*}
In modern industrial recommender systems, the focus has shifted from maximizing single interaction metrics to optimizing the overall effective Cost Per Mille (eCPM). This requires simultaneously predicting multiple heterogeneous user behaviors, ranging from shallow interactions (e.g., \textit{Click}) to deep conversion objectives (e.g., \textit{Payment}). {Multi-task learning (MTL) enables efficient multi-target prediction in recommendation by sharing knowledge across tasks, which improves performance and lowers computational cost.} While MTL architectures like MMoE \cite{mmoe} and AITM \cite{aitm} are standard for representation learning, they typically rely on a rigid \textit{inductive bias}: assuming  universal dependency strengths within the conversion funnel across all traffic.

However, this assumption overlooks the \textit{structural heterogeneity} inherent in diverse industrial applications, that is, the systematic differences in dependency patterns across conversion tasks. While the physical causal order (e.g., \textit{Click} precedes \textit{Payment}) is necessary, the \textit{relevance intensity} between these nodes varies significantly. For instance, heavy games follow a strong causal path ($Click \to Download \to Pay$), whereas lightweight tools often exhibit weaker dependencies where users skip intermediate steps ($Click \to Pay$). Enforcing a monolithic topology on such divergent instances limits model expressiveness. Although recent approaches like PMTRec \cite{pmtrec} attempt to personalize optimization weights, they fail to account for these explicit structural variations between task nodes. Furthermore, hierarchical message passing along fixed chains often leads to \textit{cumulative signal attenuation} in deep dependency paths, which significantly degrades signals for sparse downstream tasks. To address these limitations, we propose Personalized Task Dependency Graph (\textbf{PTDG}), which treats task dependency as a dynamic, learnable topology. Our main contributions are as follows:

\begin{itemize}
    \item \textbf{Dynamic Topology Learning:} PTDG utilizes a low-rank graph generation mechanism to adaptively construct {instance-specific} dependency intensities, effectively modeling heterogeneous correlations within causal constraints.
    \item \textbf{Controllable Propagation:} To mitigate signal erosion, we design a GCN-based message passing module with hard causal masking, which establishes adaptive shortcuts based on relevance.
    \item \textbf{Adaptive Optimization:} We introduce an Adaptive Progressive Masking (APM) strategy. Unlike numerical optimization methods, APM decouples parameters based on label density to resolve gradient conflicts. This approach structurally resolves the gradient conflict while reducing computational complexity.
\end{itemize}

\section{Related Work}

{Research on Multi-Task Learning (MTL) for recommendation has primarily focused on mitigating negative transfer. Architectures such as MMoE \cite{mmoe}, PLE \cite{ple}, and AITM \cite{aitm} employ gating networks to disentangle shared and specific representations. Recent work PaDiRec\cite{padirec} employs a diffusion process-based adapter to generate corresponding model parameters conditioned on dynamic task preference weights. Optimization strategies such as GradNorm\cite{gradnorm}, PMTRec \cite{pmtrec} and MoCoGrad \cite{mocograd} address gradient conflicts through re-weighting and momentum calibration, respectively. However, these methods predominantly optimize the numerical loss landscape, treating tasks as parallel objectives or using globally-shared dependency patterns, thereby neglecting the structural heterogeneity across different industrial applications.

Given the sequential nature of user actions, modeling task dependencies has become essential. ESMM \cite{esmm} and ESCM2 \cite{escm2} address sample selection bias by modeling conversion rate over the entire space. AITM \cite{aitm} introduces an information transfer module along a fixed conversion chain. HTLNet \cite{htlnet} investigates hierarchical cascading paths but enforces fixed level assignments, limiting its ability to capture instance-level variation. CSRL \cite{csrl} attempts to learn causal structures, but its iterative search incurs prohibitive costs ($\mathcal{O}(T^2)$ or higher), rendering it incompatible with industrial latency constraints. MIT \cite{mit} explores cross-task correlations but relies on globally-shared task relationships. In contrast, PTDG learns instance-specific, low-rank dependency graphs via efficient GCNs \cite{gcn, lightgcn,Continual_GCN}, balancing structural flexibility with the latency constraints of industrial systems.}

\section{Method}
As illustrated in Figure \ref{structure}, the architecture consists of three main components: (1) Personalized Topology Learning, which constructs task-to-task dependency graphs using low-rank factorization; (2) Adaptive Progressive Masking (APM), which structurally decouples parameter sharing to mitigate gradient conflicts; and (3) Causally Optimized Message Passing, which employs GCNs on hard causal masked dependency graphs with shortcut connections to prevent semantic backflow (e.g., \textit{Pay} $\to$ \textit{Click}) and signal erosion.

\subsection{Problem Formulation}
We formulate multi-objective conversion prediction as a supervised learning task. Let $\mathcal{D} = \{(d_1, Y_1), \dots, (d_N, Y_N)\}$ denote the training dataset, where each data point $d_n$ is a triplet $(f_i, f_u, f_s)$ representing item, user, and scenario features. The objective is to learn a mapping function $\Phi: \mathcal{X} \to \mathcal{Y}$, which estimates the probabilities $Y_n = \{y_n^{(1)}, y_n^{(2)}, \dots, y_n^{(T)}\}$ for $T$ distinct conversion goals.

\subsection{Personalized Task Dependency Graph}
Unlike approaches that enforce static dependency strengths, PTDG learns an item-specific graph structure within physical constraints.

\textbf{Item-Specific Dependency Graph.} To prevent overfitting on noise-prone industrial interactions and reduce parameter complexity, we adopt the low-rank approximation. The item-specific adjacency matrix $\mathbf{A}_{item}$ is defined as:
\begin{equation}
\mathbf{A}_{item} = \mathbf{Z}_i^L (\mathbf{Z}_i^R)^\top
\end{equation}
where $\mathbf{Z}_i^L, \mathbf{Z}_i^R \in \mathbb{R}^{T \times q}$ ($q < \frac{T}{2}$) are latent factors derived from the $i$-th item embeddings (extracted by two backbone models such as DeepFM \cite{deepfm} or MLP).

\textbf{Personalized Bias Injection.} To capture context-specific dependencies, we dynamically modulate the graph structure. Specifically, we generate a personalization bias vector $\mathbf{\gamma}_{u,s} \in \mathbb{R}^{q}$ from features of user $u$ and scenario $s$ using an MLP. This bias is then broadcast across all $T$ tasks to modulate the item-specific latent factors:
\begin{equation}
\mathbf{A}_{dyn} = \mathbf{Z}_{i,u,s}^L (\mathbf{Z}_i^R)^\top, \quad \mathbf{Z}_{i,u,s}^L = \mathbf{Z}_{i}^L \odot (\mathbf{1}_{T} \otimes \mathbf{\gamma}_{u,s}^\top)
\end{equation}
where $\mathbf{A}_{dyn}$ represents the personalized dynamic dependency matrix, $\mathbf{1}_{T}$ is an all-ones column vector, and $\otimes$ denotes the outer product. This mechanism provides a lightweight way to capture instance-level diversity while alleviating the overfitting problem.
\begin{table}[t]
  \caption{Statistics of Experiment Datasets.}
  \label{tab:datasets}
  \small
  \begin{tabular*}{\columnwidth}{@{\extracolsep{\fill}}lccc}
    \toprule
    Dataset & Samples & Tasks & Positive Ratio (range) \\
    \midrule
    KuaiRand1K & 8.4M & 6 & 0.02\% $\sim$ 37.9\% \\
    Industrial & 13.2M & 7 & (Confidential)\\
    \bottomrule
  \end{tabular*}
\end{table}
\textbf{Causal Pruning \& Message Passing.} We apply a hard causal mask $\mathbf{M}_{causal}$ to enforce behavior order constraints and avoid semantic backflow, followed by GCN-based propagation which operates on the original task feature matrix $\mathbf{X}$:
\begin{equation}
\mathbf{A} = \mathbf{A}_{dyn} \odot \mathbf{M}_{causal}
\end{equation}
\begin{equation}
\mathbf{H} = \sigma \left( \hat{\mathbf{D}}^{-\frac{1}{2}} (\mathbf{A} + \mathbf{I}) \hat{\mathbf{D}}^{-\frac{1}{2}} \mathbf{X} \mathbf{W} \right)
\end{equation}
where $\mathbf{H}$ represents the extracted task node feature matrix, which is used for the final CVR prediction. $\mathbf{I} \in \mathbb{R}^{T \times T}$ is an identity matrix, $\hat{\mathbf{D}}$ represents the degree matrix of $(\mathbf{A} + \mathbf{I})$, and $\mathbf{W}$ is a learnable parameter matrix. In contrast to static sequential chains, personalized dependency graph creates adaptive shortcuts between tasks without causal relation, which prevents the signal erosion caused by redundant intermediate steps in fixed hierarchies. 

We employ the extracted task node feature matrix to predict the conversion probability $\hat{y}_t$ of the user for the target $t$:
\begin{equation}
\hat{y}_t = Sigmoid\left(\text{MLP}(h_t)\right)
\end{equation}
where $h_t$ represents the feature of node $t$ for task $t$ in the extracted task node feature matrix $\mathbf{H}$.
\subsection{Optimization Objective}
The overall training objective $\mathcal{L}$ of PTDG is a weighted sum of per-task binary cross-entropy losses:
\begin{equation}
\mathcal{L} = \sum_{t=1}^{T} \beta_{t,s} \cdot \mathcal{L}_{t,s}, \quad \mathcal{L}_{t,s}=\text{BCE}(y_t, \hat{y}_{t,s})
\end{equation}
\begin{equation}
\hat{\mathcal{L}}_{t,s}=\alpha \mathcal{L}_{t,s} + (1-\alpha) \hat{\mathcal{L}}_{t,s-1}, \quad \beta_{t,s}=\frac{1}{\hat{\mathcal{L}}_{t,s}}
\end{equation}
where $\beta_{t,s}$ is the weight of task $t$ at training step $s$, and $\hat{y}_{t,s}$ is the predicted probability for task $t$ at step $s$. We obtain the task loss weights $\beta_{t,s}$ by dividing each task loss by their exponential moving average (EMA), bringing them all to a unified scale to stabilize training.

\subsection{Adaptive Progressive Masking (APM)}
Information transfer between multiple tasks may exacerbate the gradient conflict problem inherent in multi-task models. To address this, we propose a structural masking strategy that learns a task-specific mask $\mathbf{m}_t$ to select shared parameters, which is computed using the task embedding $e_t$ we add for each task.

\textbf{Sparsity-Aware Mask.} A binary mask $\mathbf{m}_t$ is generated by retaining only the top-$k_t$ elements of a learned gate:
\begin{equation}
\mathbf{m}_t^{(j)} = \mathbb{I}\left(\text{rank}(\mathbf{w}_t^{(j)}) \le k_{t}\right), \quad \mathbf{w}_t = \text{Gate}_t(\mathbf{e}_t)
\end{equation}
where $\mathbf{m}_t^{(j)}$ represents the $j$-th parameter of mask $\mathbf{m}_t$. The mask is applied to the parameters shared between tasks through the Hadamard product to obtain the parameters visible to task $t$.
\begin{table}[t]
  \caption{{Performance comparison in terms of AUC on KuaiRand dataset and industrial dataset. Values are means over 20 different random seeds. Bold indicates the best, underline the second best.}}
\centering
\renewcommand\arraystretch{1.2}
\label{tab:comparison}

\resizebox{1.0\linewidth}{!}{
\begin{tabular}{c|c|ccccccc}
 \toprule[1.2pt]
    Dataset & Task & MMoE & PLE & STEM & {MCGrad} & {PMTR} & MIT & \textbf{{PTDG}} \\
 \midrule
 \multirow{7}{*}{KuaiRand1K}
    & Click     & 0.8706 & 0.8637 & 0.8732 & 0.8743 & \underline{0.8772} & 0.8766 & \textbf{0.8801} \\
    & Like      & 0.8879 & 0.8903 & 0.8998 & 0.9009 & 0.9017 & \underline{0.9028} & \textbf{0.9085} \\
    & Follow    & 0.9033& 0.9024 & 0.9073 & \underline{0.9130} & 0.9124 & 0.9129 & \textbf{0.9193} \\
    & Comment   & 0.9051& 0.8986 & 0.9081 & 0.9072 & 0.9085 & \underline{0.9101} & \textbf{0.9172} \\
    & Forward  & 0.8984& 0.8933 & 0.9028 & 0.9011 & \underline{0.9042} & 0.9038 & \textbf{0.9097} \\
    & Hate     & 0.9041& 0.8960 & 0.9055 & 0.9066 & 0.9072 & \underline{0.9085} & \textbf{0.9139} \\
    \cmidrule(lr){2-9}
    & \textit{Avg.} & \textit{0.8949}& \textit{0.8907} & \textit{0.8995} & \textit{0.9005} & \textit{0.9019} & \textit{\underline{0.9025}} & \textbf{\textit{0.9081}} \\
    \midrule
    \multirow{8}{*}{Industrial}
    & Task 1 & 0.8410& 0.8424 & 0.8494 & \underline{0.8501} & {0.8481} & 0.8471&\textbf{{0.8527}} \\
    & Task 2 & 0.8395& 0.8433 & 0.8457 & {0.8462} & {\underline{0.8500}} & 0.8468&\textbf{{0.8511}} \\
    & Task 3 & 0.8601& 0.8379 & 0.8482 & {0.8550} & {\underline{0.8745}} & 0.8541&\textbf{{0.8872}} \\
    & Task 4 & 0.9417& \textbf{0.9471} & 0.9457 & \underline{0.9468} & {0.9462} & 0.9452&{0.9464} \\
    & Task 5 & 0.8699& 0.8678 & 0.8613 & {0.8630} & {\underline{0.8729}} & 0.8732&\textbf{{0.8751}} \\
    & Task 6 & 0.8837& 0.8796 & 0.8557 & {0.8610} & {0.8843} & \underline{0.8899}&\textbf{{0.8993}} \\
    & Task 7 & 0.8912& 0.8974 & 0.8915 & {0.8952} & {0.8967} & \underline{0.8979}&\textbf{{0.9028}} \\
    \cmidrule(lr){2-9}
    & \textit{Avg} &\textit{0.8753} & \textit{0.8736} & \textit{0.8710} & \textit{{0.8739}} & \textit{{\underline{0.8818}}}& \textit{0.8791} & \textbf{\textit{{0.8878}}} \\
 \bottomrule[1.2pt]
 \end{tabular}}
\end{table}
\textbf{Adaptive Progressive Masking.} The parameter mask rate $r_{target}^{(t)}$ is determined by the label positive rate $p_t$ of task $t$ to balance easy (dense) and hard (sparse) tasks:
\begin{equation}
r_{target}^{(t)} = r_{min} + (r_{max} - r_{min}) \cdot \left(1 - \frac{p_t - p_{min}}{p_{max} - p_{min}}\right)
\end{equation}
where $p_{min}$ and $p_{max}$ represent the minimum and maximum label positive rate,  $r_{max}$ and $r_{min}$ are hyper parameters. We also implement a linear warm-up mechanism with max warm-up step $S_{warm}$ to adjust the learning difficulty gradually based on the training step $s$ and the number of shared parameters $d$ to stabilize training:
\begin{equation}
k_{t,s} = d - d \cdot \min \left( r_{target}^{(t)}, \frac{s}{S_{warm}} \cdot r_{target}^{(t)} \right)
\end{equation}

\subsection{Complexity Analysis}
PTDG uses low-rank factorization ($O(T \times q)$) as structural regularization rather than directly learning a full adjacency matrix, which prevents overfitting to spurious correlations on sparse industrial data. The low-rank formulation ensures both adjacency calculation and GCN-based message passing remain linear in $T$. In production, PTDG increases end-to-end latency by only 8\,ms over the MMoE baseline, within serving SLA. Unlike gradient-based de-conflicting methods (e.g., PCGrad \cite{pcgrad} with $\mathcal{O}(T^2)$ projections), APM introduces negligible training cost and no inference overhead as masks are fixed after training.

\section{Experiments}
\subsection{Experimental Setup}
\textbf{Datasets.} We evaluate PTDG on KuaiRand1K (6 tasks) and an industrial dataset (7 tasks). Details are provided in Table \ref{tab:datasets}.

\textbf{Baselines.} We compare PTDG with strong industrial baselines and state-of-the-art methods:
(1) MMoE \cite{mmoe}: A widely adopted multi-gate mixture-of-experts architecture.
(2) PLE \cite{ple}: A widely used industrial standard for parameter sharing.
(3) STEM \cite{stem}: An advanced optimization-based MTL framework.
(4) MoCoGrad \cite{mocograd}: A method addressing gradient conflicts via momentum calibration.
(5) PMTRec \cite{pmtrec}: A personalized multi-task approach using gradient re-weighting.
(6) MIT \cite{mit}: A recent model focusing on cross-task heterogeneous correlation modeling.

\textbf{Implementation Details.} We use Adam as our optimizer with a learning rate of 1e-3, a dropout rate of 0.1, and a batch size of 2048. In all experiments, we repeat each setting 20 times using 20 different random seeds. The rank $q$ of item-specific latent factors is set to 2. The parameter $\alpha$ in the EMA-based multi-task loss weight optimization is set to 0.1. The maximum and minimum mask rates $r_{max}$ and $r_{min}$ of APM were set to 0.7 and 0.2, respectively. The warm-up process in APM was conducted with $p_{min}=500$ and $p_{max}=2000$. All baseline method parameters were kept consistent with those in the original paper. All experiments were conducted on a single NVIDIA Tesla V100 GPU.

\subsection{Performance Comparison (RQ1)}
\textbf{Comparison with SOTAs.}
We conducted 20 runs using 20 different random seeds. As shown in Table~\ref{tab:comparison}, PTDG achieves the highest average AUC on both datasets.
PTDG surpasses MIT~\cite{mit} by 1.05\% on Task 6---the key deep conversion task---because, unlike MIT's rigid global correlations, it adapts task relationships per instance, enabling GCN-based message passing that preserves signal strength across deeper tasks.
PTDG also beats MoCoGrad~\cite{mocograd} by 1.59\% on average. Unlike MoCoGrad's post-hoc momentum tuning, our structural APM directly resolves gradient conflicts at the parameter level, yielding more effective and efficient multi-task optimization. 

\begin{table}[t]
  \small
  \centering
  \caption{Ablation Study on Industrial Dataset.}
  \label{tab:ablation}
  \begin{tabular*}{\columnwidth}{@{\extracolsep{\fill}}clc}
    \toprule
    \quad Exp. & Variant Model & Avg AUC  \\
    \midrule
    \quad 1 & \textbf{PTDG (Full)} & \textbf{0.8878}  \\
    \quad 2 & w/o Personalization ($\mathbf{\gamma}_{u,s}$) & 0.8771 \\
    \quad 3 & w/o GCN (use Concat) & 0.8796 \\
    \quad 4 & w/o Causal Mask & 0.8814 \\
    \quad 5 & w/o APM (Adaptive Masking) & 0.8826 \\
    \quad 6 & w/o Progressive Warmup & 0.8841 \\
    \bottomrule
  \end{tabular*}
\end{table}

\textbf{Performance on Deep Tasks.} 
{A critical observation is PTDG's strong performance on sparse, deep-conversion objectives. On dense, shallow tasks (e.g., Task 4, click), PTDG achieves comparable but not significantly better performance, which is expected as signal erosion primarily affects sparse deep-funnel objectives. On harder tasks, however, PTDG significantly outperforms state-of-the-art methods. For instance, on the sparse Industrial Task 3, PTDG surpasses PMTRec by 1.45\%. Similarly, on the follow task in KuaiRand1K, it exceeds our online baseline model MMoE by 1.77\%. This confirms that our personalized graph topology effectively mitigates signal erosion for deep nodes without compromising head-task performance.}

\subsection{Ablation Study (RQ2)}
To verify the contribution of each component, we conduct an ablation study on the Industrial Dataset (Table~\ref{tab:ablation}). Removing instance-level dependency graph decreases Avg AUC by 1.21\%, confirming that personalizing task dependencies is effective for capturing heterogeneous multi-task relationships. Replacing the GCN with simple concatenation lowers AUC by 0.93\%, verifying that GCN-based message passing effectively preserves signal strength across tasks compared to passing through the static conversion funnel. Disabling adaptive masking (w/o APM) reduces performance by 0.59\%, underscoring its effectiveness in mitigating gradient conflicts.

\subsection{Sensitivity Analysis}
We vary the low-rank dimension $q$ from 1 to 7, performance peaks at $q=2$ and degrades with larger $q$, confirming that low-rank approximation effectively controls overfitting while using fewer parameters than a full adjacency matrix. Empirically, $q \approx T/3$ works well across our settings. Formal scaling analysis is left to future work. We also conduct masking rate experiments on tasks with maximum and minimum masking rates ranging from 0 to 0.9 in the APM strategy. The experimental results show that the model achieves the best performance when set to \(r_{min} = 0.2\) for high positive sample rate tasks and \(r_{max} = 0.7\) for low positive sample rate tasks.

\subsection{Online A/B Testing}
We deployed PTDG as the ranking model of a mainstream app distribution platform, serving over 100 million daily active users, for a two-week online A/B test. 
The baseline model was the currently deployed MMoE. This is the same MMoE architecture included in our offline comparison (Table~\ref{tab:comparison}). Traffic was split into two buckets: 10\% for the base bucket and 10\% for the experimental bucket. 
The remaining 80\% of traffic continued to use the original production model to ensure service stability. 
{To ensure statistical significance, we computed daily aggregated metrics for each bucket and performed two‑sample t‑tests with a significance threshold of p < 0.05.}
Results show that PTDG achieved statistically significant improvements on core business metrics, with a +1.2\% lift in CVR and +1.9\% lift in eCPM. This demonstrates the industrial viability of PTDG as a multi-task learning framework that delivers consistent gains in real-world recommender systems.

\section{Conclusion}
This work addresses signal erosion in industrial multi-task recommendation through the proposed PTDG framework. Unlike static funnel-based approaches, PTDG employs a low-rank dynamic graph learning method to model instance-level dependencies under strict latency constraints. By combining causal message passing with structural parameter decoupling, the model mitigates signal attenuation for deep-conversion objectives. Experiments demonstrate that PTDG achieves a favorable trade-off, improving performance on sparse, high-value tasks (up to 1.45\% lift) while maintaining accuracy on dense interactions.  Future work will explore scaling to larger task sets and extending the framework to cross-domain settings.

\section*{GenAI Usage Disclosure}
Generative AI tools assisted with language editing and code debugging. All experimental results were produced by author-written scripts and verified against saved artifacts.

\bibliographystyle{ACM-Reference-Format}
\balance
\bibliography{sample-base}

\end{document}